\input{psfig.sty}
\documentclass[longauth]{aa}
\usepackage{amssymb}
\usepackage{txfonts}
\usepackage{natbib}

\newcommand\sw{{\it Swift}}
\newcommand\pc{{\it pc}}
\newcommand\wt{{\it wt}}
\newcommand\im{{\it im}}
\begin{document}
\title{GRB051210: \sw~detection of a short gamma ray burst}
   \author{V. La Parola \inst{1},
V.~Mangano\inst{1}, 
D.~Fox\inst{2},
B.~Zhang\inst{3}, 
H.A.~Krimm\inst{4,5},
G.~Cusumano\inst{1},
T.~Mineo\inst{1}, 
D.~Burrows\inst{2},
S.~Barthelmy\inst{4},
S.~Campana\inst{6}, 
M.~Capalbi\inst{7},
G.~Chincarini\inst{6,8},
N.~Gehrels\inst{4},
P.~Giommi\inst{7},
F.E.~Marshall\inst{4},
P. M\'esz\'aros\inst{2,9},
A.~Moretti\inst{6},
P.T.~O'Brien \inst{10},
D.M.~Palmer\inst{11},
M.~Perri\inst{7},
P.~Romano\inst{6},
G.~Tagliaferri\inst{5}.
          }
\offprints{V. La Parola}

\institute{INAF- Istituto di Astrofisica Spaziale e Fisica Cosmica di Palermo
	   \email{laparola@ifc.inaf.it}
      \and
          Department of Astronomy and Astrophysics, Pennsylvania State University, 525 Davey Laboratory, University Park, PA 16802, USA
      \and
	  Department of Physics, University of Nevada, 4505 Maryland Parkway, Las Vegas, NV 89154-4002, USA
      \and
          NASA Goddard Space Flight Center, Greenbelt, MD 20771, USA
      \and
          Universities Space Research Association 10211 Wincopin Circle, Suite 500 Columbia, Maryland 21044, USA
      \and
	  INAF -- Osservatorio Astronomico di Brera, Via Bianchi 46, 23807 Merate, Italy
      \and
	  ASI Science Data Center, via Galileo Galilei, 00044 Frascati, Italy
      \and
	  Universit\`a degli studi di Milano-Bicocca, Dipartimento di Fisica, Piazza delle Scienze 3, I-20126 Milan, Italy
      \and
          Department of Physics, Pennsylvania State University, PA 16802, USA
      \and
          Department of Physics and Astronomy, University of Leicester, University Road, Leicester, LE1 7RH, UK
      \and
          Los Alamos National Laboratory, MS B244, NM 87545, U.S.A.
     	  }

   \date{Received xxxx, 2006/ accepted xxxx, 2006}
\abstract
   {The short/hard GRB051210 was detected and located by the \sw-BAT instrument
   and rapidly pointed towards by the narrow field instrumens. The XRT was able
   to observe a bright X-ray afterglow, one of the few ever observed for this
   class of bursts.}
   {We present the analysis of the prompt and afterglow emission of this event}
   {The BAT spectrum is a power-law with photon index $1.0\pm0.3$.
   The X-ray light curve decays with slope $-2.58\pm0.11$ and shows a small flare in the
   early phases. The spectrum can be described with a power law with photon
   index $1.54\pm0.16$ and absorption $(7.5_{-3.2}^{+4.3})\times10^{20}$ cm$^{-2}$}
   {We find that the X-ray emission is consistent with the hypothesis that we
   are observing the curvature effect of a GRB occurred in a low density medium,
   with no detectable afterglow. We estimate the density of the circumburst
   medium to be lower than $4\times10^{-3}$ cm$^{-3}$. We also discuss different
   hypothesis on the possible origin of the flare.}
   {}
   \maketitle

\section{Introduction}
It has long been known that the T$_{90}$ duration and hardness ratio of
the population of gamma-ray bursts (GRBs) show a bimodal distribution, where
two classes can be identified: long GRBs, with duration longer than 2 seconds 
and short GRBs lasting less than 2 seconds and showing a harder spectrum 
\citep{mazets,norris,kou}. While long GRBs have been studied in good detail and 
their origin is now
established in the explosion of massive stars leading to very energetic core
collapse supernovae (e.g. \citealp{woos,bloom,hjo}), very little was known about 
short GRBs, mainly because of the difficulty to localize them with good 
accuracy. A good progenitor candidate for short GRBs has been identified in the 
merger of two compact objects in a tight binary (e.g. \citealp{eichler})\\
Thanks to the rapid repointing capability of the \sw~satellite \citep{gehr}, 
that allows for an accurate localization of the afterglow within few minutes 
from the burst onset, this gap is now being filled, and up to now seven short 
GRBs have been localized and their afterglow detected (GRB050509B:
\citealp{gehrnat,bloom05}; GRB050709: \citealp{villa,fox}, GRB050724:
\citealp{barthe,campana}; GRB050813: \citealp{fox2}; GRB051210 \citealp{4315},
GRB051221A: \citealp{4363}; GRB060121: \citealp{4550}). The association of two of 
these events (GRB050509B and GRB050724) with late-type
galaxies and the amount of energy involved (significantly lower than for long
GRBs) support the hypothesis that short/hard bursts are the product of the 
merger of two compact objects in a binary system. The lack of any supernova 
signature in the identified host galaxies confirms this idea. On the other 
hand, the association of GRB050709 and GRB051221A with star forming galaxies is 
not at variance with this hypothesis, and merely extends the range of possible 
lifetimes of the progenitor system, which can be located both in early-type, 
old population galaxies, and in star-forming galaxies.\\
GRB051210 triggered the \sw-BAT instrument \citep{barth05} on 
December 12 2005 at 05:46:21 UT \citep{4315}. The BAT position calculated 
on-board was RA=22h 00m 47s, Dec=-57\degr 38\arcmin 01\arcsec (J2000), with a
90\% uncertainty of 3\arcmin. 
The burst was classified as short after the on-ground analysis of the 
BAT data \citep{4318,4321} \citet{4321} report a spectral lag of
$-0.0010^{+0.01500}_{-0.0170 sec}$, typical of short GRBs \citet{nor2}. The 
spacecraft slewed immediately and the XRT 
\citep{bur05} and UVOT \citep{rom05} began observing the field 79.2 and 70 s 
after the trigger, respectively. The XRT onboard centroiding procedure found a 
bright fading uncatalogued X-ray source. No source was 
detected in any of the UVOT filters \citep{4331}. Two sources were detected
within the XRT error circle by the 6.5m Clay/Magellan using the LDSS3 
instrument: \citet{4330} report a clear detection of an apparently extended 
(north by north-east) source 2\arcsec.9 from the XRT position and a second 
marginal detection 1\arcsec.1 from the XRT position.\\
In this paper we report on the analysis of the prompt and afterglow emission of
GRB051210 as observed by the \sw~ X-ray instruments. Sections~\ref{bat} and 
\ref{xrt} describe the observations, the data reduction and the analysis of the BAT and 
XRT data respectively. The results are discussed in Sect.~\ref{disc}.
  
\section{BAT observation}
\label{bat}
\begin{figure}[th]
\centerline{\psfig{figure=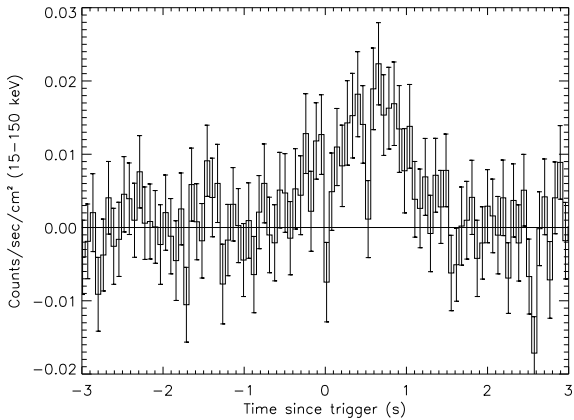,width=8cm}}
\caption{BAT light curve of the prompt emission of GRB051210. Each time bin is
64 ms long.}
\label{blc}
\end{figure}

The BAT event data were re-analyzed using the standard BAT analysis
software included in the HEASOFT distribution (v. 6.0.3), as described in the 
Swift BAT Ground Analysis Software Manual \citep{krimm}, that incorporates 
post-launch updates to the BAT response and to the effective area and includes 
the systematic error vector to be applied to the spectrum. The light curve
showed a double peak with T$_{90}$=$1.27\pm0.05$ sec 
(Fig.~\ref{blc}). The average spectrum can be described with a power law with 
$\Gamma=1.1\pm0.3$, with a total fluence of $(8.1\pm1.4)\times10^{-8}$ erg/cm$^2$ in 
the 15 150 keV band. A fit with a Band model is unconstrained. 
\subsection{Data reduction}
The XRT is designed to perform observations 
in different read-out modes (see \citealp{hill} for a detailed description of
the XRT modes), switching automatically among them according to the
count rate level of the source, in order to optimize the collected information 
and minimize the pile-up in the data. At present, the Imaging (\im), Windowed
Timing (\wt) and Photon Counting (\pc) modes are fully operating.\\
XRT began observing the field of GRB051210 in auto state and went through the 
correct sequence of read-out modes. The first \im~frame (lasting 2.5 sec) 
allowed for the on-board localization of the burst \citep{4315}. This was 
followed by 85.1 sec in \wt~mode and then by 37.0 ksec in \pc~mode. \\
They were calibrated, filtered and screened using the XRTDAS (v.2.3) 
software package to produce cleaned photon list files.\\
The position of the source was recalculated on ground using the task {\sc
xrtcentroid} on the \pc~image and applying the boresight correction through 
the updated TELDEF file 
provided by the Swift Science Data Center \citep{4313}. The refined XRT
coordinates are RA=+22h 00m 41.3s, Dec=-57\degr 36\arcmin 
48\arcsec.2 (J2000), with 4\arcsec.2 uncertainty \citep{4320}. \\
The photons for the timing and spectral analysis were extracted from a region 
with 20 and 30
pixels radii for \wt~and \pc~data, respectively. In order to account for the 
pile-up in the \pc~data, photons from a radius of 2.5 pixels around the source 
centroid were excluded from the analysis, and the remaining photons were then
corrected by the fraction of point spread function lost. The presence of a hot 
column crossing the source close to the centroid was accounted for, both in 
the \wt~and \pc~data using a correction factor derived from the ratio of the 
effective areas calculated for the same region with and without the hot column. 
The count rate of the source during the \im~ frame was obtained integrating the
DN above the background in a 30 pixel radius circle and then following the
procedure described in \citet{goad}. \\
\begin{figure}[th]
\centerline{\psfig{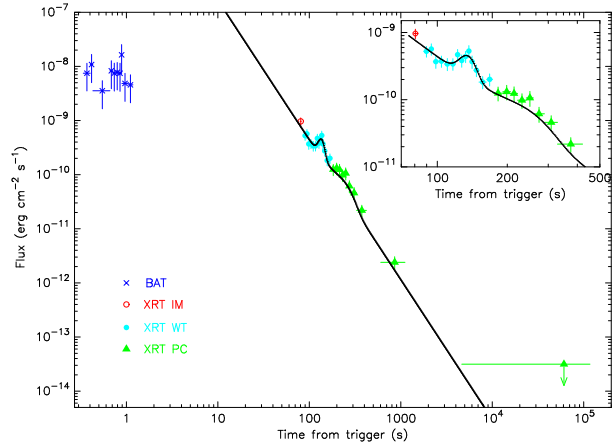}}
\caption{XRT light curve decay of GRB\,051210. The XRT count rate (0.2-10 keV)
was converted into flux units by applying a conversion factor derived 
from the spectral analysis. The solid line represents the best fit model to the
XRT data. The BAT light curve was extrapolated into the XRT energy band 
by converting the BAT count rate with the factor derived from the BAT 
spectral parameters. }
\label{lc}
\end{figure}
\section{XRT observation}
\label{xrt}
\subsection{Data analysis}
\begin{figure}[th]
\centerline{\psfig{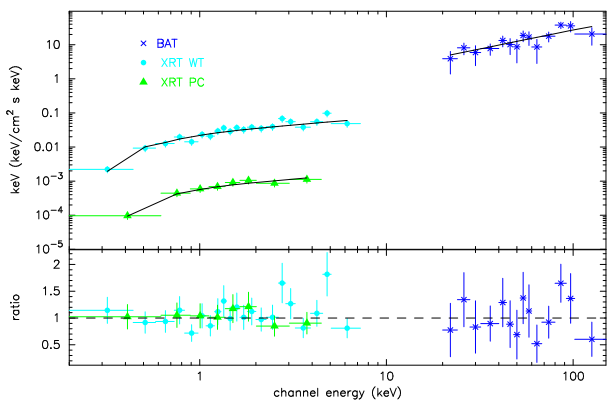}}
\caption{XRT 0.2-10 keV and BAT 15-150 keV energy spectra of GRB\,051210,
 with the best-fit absorbed power law model and residuals. The BAT and XRT data
 are not simultaneous.}
\label{spec}
\end{figure}

Figure~\ref{lc} shows the BAT and XRT light curve of GRB051210. The BAT data 
(originally in the 15-150  keV band) were extrapolated into the XRT energy band 
(0.2-10 keV) and the source observed count rates converted into flux using the 
appropriate conversion factor derived from the spectral analysis. The X-ray
afterglow is detected only in the first 1000 s after the onset.
The XRT light curve can be modelled with a single power law with decay index
$2.58\pm0.11$. The flare in the early afterglow is well described by a Gaussian
centered at $134\pm2$ sec with $\sigma=10\pm2$ s. The inclusion of this component in the fit
improves the $\chi^2_{red}$(d.o.f.) from 2.17(23) to 1.14(20), with F-test
chance probability of $1\times10^{-3}$. A second Gaussian centered at $216\pm44$
sec yields a further marginal improvement to the fit, with  F-test
chance probability of $4\times10^{-2}$.\\
In order to compare the decay before and after the flares, we made an estimate 
of the decay slope before the flares by comparing the photon arrival times in 
the first 26 seconds of the \wt~mode observation with power laws with different 
slopes through a Kolmogorov-Smirnov test. We get 
$\alpha_{wt}=2.35^{+0.20}_{-0:30}$,
where the best value is the one that maximizes the probability
that the photon arrival times follow a power law distribution,
and the quoted errors define the interval of the slope values for
which the test provides an ``acceptance'' probability higher than
90\%. This slope is consistent both with the average decay and
with the late time light curve.\\
The XRT spectrum (Fig~\ref{spec}, Table~\ref{spec}) is relatively hard and 
can be well described with an absorbed power law ($\Gamma=1.54\pm0.16$) with 
absorption slightly in excess with respect to the Galactic line of sight value 
($2.2\times10^{20}$ cm$^{-2}$, \citealp{nh}). 
All quoted errors are at 90\% confidence level.

\begin{table}
\caption{Spectral fit results}             
\label{spectab}      
\centering                          
\begin{tabular}{c c c}        
\hline\hline                 
                       & BAT         & XRT  \\    
\hline                        
N$_h$ (cm$^{-2}$)      & --          & $(7.5_{-3.2}^{+4.3})\times10^{20}$\\
$\Gamma$               & $1.1\pm0.3$ & $1.54\pm0.16$\\      
$\chi^2_{red}$ (d.o.f.)& 1.0 (57)      & 0.78 (25) \\
\hline                                   
\end{tabular}
\end{table}

\section{Discussion}
\label{disc}
GRB051210 is one of the few short GRBs for which we have an X-ray detection and
an accurate location. Its X-ray light curve decays rapidly and the X-ray 
counterpart is not detectable anymore after $\sim$1 ks. The X-ray light curve 
decays as a power law with slope $\alpha = 2.57\pm0.11$. Superimposed on it we 
detect a flare peaking at T+134 s, and a less significant bump at T+216 s. The 
BAT and XRT spectra can be described with a power law, with photon indexes $1.0
\pm0.3$ and $1.54\pm0.16$ respectively. \\
The shape of the light curve is very similar to that of GRB050421 \citep{godet}.
The rapid fading of the source and the 
lack of any flattening in the light curve after the first steep decay may 
indicate that the GRB occurred in an extremely low density medium (naked GRB, 
\citealp{kumar,page}) where the radiation emitted by the forward shock, generated by 
the impact of the initial shock front with the surrounding interstellar medium,
is expected to be undetectable. The steep decay of the X-ray emission is 
fully consistent with the hypothesis that we are observing a low energy tail 
of the prompt emission from an internal shock through the so-called  
curvature effect  \citep{kumar,dermer,zhang}: the radiation observed as the tail of 
a peak is expected to be the off-axis emission of the shocked surface arriving 
at the observer at later times, and would decay as
$t^{-\alpha}=t^{-(\Gamma+1)}$, where $\Gamma$ is the photon index of the GRB
emission. In the case of GRB051210 we get $\alpha=\Gamma+1=2.54$, in very good 
agreement with the observed slope ($2.57\pm0.11$). However, the extrapolation 
of the XRT light curve back to the burst onset does not match the BAT points 
by a few decades. This could well be due to the fact that the XRT light curve 
is in fact the tail of a flare peaked at the time before the XRT 
observation and too weak to be detected by the BAT. Within such an
interpretation, the zero time point of the rapid decay component
should be shifted to the beginning of the rising segment of the
relevant flare \citep{zhang,liang}, which marks the reactivation of the
central engine. Visual inspection of the lightcurve suggests that this
time is about $~$10 s. We then re-fit for the decay index after
such a shift. The temporal decaying slope is changed
$2.59\pm0.08$, which is still consistent with the theoretical
predicted value $\alpha=\Gamma+1=2.54$. 

If we are seeing only the tail of the prompt emission, and the afterglow is not 
detectable, we can derive an estimation for the density of the interstellar 
medium $n$ in the vicinity of the burst. Assuming $F_{LIM}=1\times10^{-14}$ erg
cm$^{-2}$ s$^{-1}$ to be the limiting flux in our observation, and considering 
that we have no detectable emission at T$_0+10^4$, we can infer $n$ from the 
expression of the expected afterglow flux according to the standard afterglow 
models \citep{sari,spn}. We assume as z the average redshift measured for short 
GRBs up to now (0.35), and estimate the energy of the afterglow ($E_a$) from 
the 1-1000 keV fluence of the prompt emission, as indicated by \citep{frail}. 
We account for the decay of the afterglow at a time T$_0+10^4$, assuming an 
electron index p=2.2 and we assume to be at a frequency between the peak 
frequency  $\nu_m$ and the cooling frequency $\nu_c$. With these assumptions, 
we get $n<4\times10^{-3}$, that confirms the trend put in evidence by
\citet{sode}, that short GRBs tend to be in low-density environments. 
We note that the estimate of $n$ is subject to uncertainties of $z$,
$E_a$ and $p$ (e.g. it increases with $z$ and $p$ but decreases with
$E_a$). In any case, the inferred density is lower than the
typical values inferred from long GRBs \citep{panaitescu}.
 
Flaring activity has been previously observed in other short GRBs: GRB050709 
\citep{fox}, that shows a flare between 25 and 130 s and a late flare 
at about 16 d, and GRB050724 \citep{barthe,campana}, that shows at
least four flares.
Interestingly, the epoch of 
the first flare is $\sim100$ s for both of those two events and for GRB051210. 
Moreover, evidence for X-ray emission at this timescale has been reported in 
the stacked light curves of several BATSE short GRBs \citep{lazzati}. 
Delayed activity from the inner engine has been generally invoked to
interpret flares \citep{bursci,zhang,romano,falcone}. This is less
problematic for long duration GRBs since in the collapsar scenario,
there is a large reservoir of the fuel and the fragmentation or
gravity instability in the collapsing star may form clumps that are
accreted at different times, leading to delayed X-ray flares
\citep{king}. However, this cannot be applied to short 
GRBs, if we accept the hypothesis that they originate in the merger of two 
compact objects in a binary system (NS-NS or NS-BH), a scenario
supported by the recent observations.
Hydrodynamical simulations suggest that the central engine activity of
merger events cannot last more than a few seconds \citep{davies}. 
\citet{perna} suggest a common origin for flares in long and
short GRBs: some kind of instability (likely gravitational instability) can 
lead to the fragmentation of the rapidly accreting accretion disk that forms 
after the GRB (both in the collapsar and in the merger model), creating blobs 
of material whose infall into the central object produces the observed flares. 
\citet{proga} suggest that magnetic fields may build up near the black
hole and form a magnetic barrier that temporarily block the accretion
flow. The interplay between the magnetic barrier and the accretion
flow can turn on and off the accretion episodes, leading to erratic
X-ray flares at late epochs. This mechanism also applies to both long
and short GRBs. \citet{dai} propose that the postmerger product for
the NS-NS system may well be a massive neutron star if the neutron
star equation of state is stiff enough. The differential rotation of
the neutron star would lead to windup of magnetic fields, leading to
magnetic reconnection events that power X-ray flares. These scenarios
are all consistent with a magnetic origin of the flares based on
energetics arguments \citep{fan}. Finally, an
alternative hypothesis has been formulated by \citet{mcfay}: the interaction 
of the GRB outflow with a noncompact stellar object is suggested as a 
natural explanation for a flare after the burst. 
This model is restricted to interpret only one flare (and therefore cannot
be applied to GRB 050724 but may be relevant for GRB 051210) and the
outflow is required to be not collimated.
In any case, the presence of a flare at a similar epoch (around 100 s)
in three out of five short bursts observed by now, if also confirmed
by future short GRBs, is a point that is worth investigating and calls
for better understanding.
  
This work is supported at INAF by funding from ASI, at Penn State by NASA and 
at the University of Leicester by the Particle Physics and Astronomy Research 
Council. We gratefully acknowledge the contribution of dozens of members of 
the XRT team at OAB, PSU, UL, GSFC, ASDC and our subcontractors, who 
helped make this instrument possible.

\end{document}